\documentstyle[12pt]{article}
\begin{document}
\baselineskip = 20pt

\begin{titlepage} 
\bigskip
\hskip 3.7in\vbox{\baselineskip12pt
\hbox{SWAT-213}
\hbox{hep-th/9901028}}
\bigskip\bigskip\bigskip\bigskip

\centerline{\large \bf Non-Perturbative Structure in Heterotic}
\centerline{\large \bf Strings from Dual F--Theory Models}
\bigskip\bigskip
\bigskip\bigskip

\centerline{D\'{o}nal O'Driscoll} 
\bigskip\bigskip

\centerline{Department of Physics, University of Wales Swansea }
\centerline{Singleton Park, Swansea, SA2 8PP, UK}
\centerline{{\em pydan@swansea.ac.uk}}
\bigskip\bigskip

\begin{abstract}
\baselineskip=16pt
We examine how to construct explicit heterotic string models dual to F--theory in eight dimensions. In 
doing so we learn about where the moduli spaces of the two theories overlap, and how non--perturbative 
features leave their trace on a purely perturbative level. We also briefly look at the relationship with NS9--branes
\end{abstract}
\end{titlepage}

\section{Introduction}
Over the last few years the idea of string duality has lead to much
greater understanding of the non--perturbative features of string
theory, to the extent that we can now visualize the various string
theories as being different points in a larger moduli space. Most
notably we have learned about the interplay of geometric features in
compactification, especially those based on the rich area of
Calabi-Yau manifolds.

In particular we have learned to relate strongly coupled type IIB
superstrings in a background of 24 7--branes to heterotic string
theory through elliptically fibered K3 surfaces \cite{v,a,
brz}. However, to date the majority of this work has been very
mathematical \cite{fmw} in nature with little attention being paid
to the explicit duality map. In this letter we will address this
issue. There are several well known and phenomenologically interesting
methods of constructing heterotic theories in less than ten dimensions
that have been known for some time, viz.: covariant lattices
\cite{covlattice}, free
fermionic constructions
\cite{freefermion}, and asymmetric orbifolds \cite{asyorbifolds}. 
It can be shown
that these are all essentially equivalent \cite{dunbar}, with each method having its
own benefits and drawbacks.

Nevertheless, it is not trivial to determine a map between F--theory
and the heterotic string compactifications directly because it is not
known to what extent the moduli spaces overlap. For example, F--theory
on K3 has a fixed supersymmetry \cite{asp}, however it is relatively easy to
construct heterotic models in eight dimensions with less
supersymmetry. Since F--theory is non--perturbative understanding the
map should provide an interesting relationship between perturbative
and non--perturbative aspects of the heterotic string.

In section 2, the basic results needed from
F--theory on how to read off gauge groups with the necessary
substructure indicating non--perturbative contributions are outlined. 
In section 3, we 
look at the prescription to recover the purely perturbative heterotic
theory and discuss how to construct the dictionary of the two via the
moduli space of Type I theory. In section 4, we conclude the paper with a brief look at how we can relate the work of this paper to the recent work being done on NS9--branes, which are problematic in string theory but appear to be required by duality arguements.

\section{F--theory}
F--theory \cite{v} is not a string theory per se, though attempts have been made to define it as a 12 
dimensional theory with two time dimensions. A much more satisfactory approach is to consider it as 
Type IIB superstring compacted on a sphere (the complex projective surface) in the background of 
twenty four 7--branes.  Type IIB superstring theory in ten dimensions has an $SL(2, {\bf Z})$ 
self--duality and hence has an associated torus. When this torus is fibered over the sphere in the brane 
background an elliptically fibered K3 surface is formed, the properties of which are well known. The 
moduli space is \cite{asp}
\begin{equation}
{\cal M}= SO(2,18;{\bf Z})\backslash SO(2,18)/SO(2) \times SO(18)
\end{equation}
and is the same as that of a heterotic string compactified on $T^2$ using a Narain lattice.

Since the K3 surface has an elliptic structure its 
singularity structure can be  easily read off
from the Weierstrass equation.
These singularities have been classified by Kodaira in a way
corresponding to the A--D--E classification of Lie algebras
\cite{bsv}. It is standard to accept this correspondence as being exact,
i.e. the singularity type corresponds to a gauge group of the same Lie
algebra type in the physical theory. However, Witten \cite{w} has
shown that this is not necessarily true, though how this works from a
heterotic point of view is not yet fully understood. Other algebras
can also be constructed using various configurations of mutually
non--local 7--branes but as they do not coalesce to a single point
they are not of interest here.

Recently there has been much work done in understanding how the gauge
groups arise from the K3 surface through the theory of string networks
\cite{networks}. The singularities of the K3 surface correspond to the
positions of the twenty four 7--branes forming the background in
F--theory. We know from work on the related theories of D--branes
\cite{cjp} that perturbatively there should be only sixteen
D7--branes. By using Seiberg--Witten theory, Sen \cite{s} showed using 
F--theory, that the orientifold planes could be formed out of 7--branes which 
have different charges with respect to the Ramond and Neveu--Schwarz sectors. 
In references \cite{gz, j} it is shown how to combine these mutually
non-local branes to provide the states needed to fill out the gauge
groups corresponding to the respective singularities on the K3 surface.

The 7--branes are classified by the RR and NS--NS charges, $[p,q]$,
they carry; 7--branes with different values of $[p,q]$ are said to be
mutually non--local. From \cite{j} we will be only concerned with
three types of mutually non--local 7--branes, denoted $A, B, C$. If we
have $n$ 7--branes of the same type at a singularity then the
associated gauge algebra is $su(n)$. If there are different types of
branes at a singularity, $A^{n_a}B^{n_b}C^{n_c}$, the associated
algebra is $su(n_a) \otimes su(n_b) \otimes su(n_c)$. This is actually
a maximal subalgebra of a larger simply laced algebra since extra
massless BPS states also appear in representations of the subalgebra
and fill out the adjoint of the larger group in a manner analogous to
free fermion constructions \cite{klt}. A
$D$ type singularity corresponds to a 7--brane configuration of the form
$A^{n}BC$; the subalgebra is $su(n) \otimes u(1)$ which is enhanced to
a $D_n$ algebra. Similarly an $E$ type singularity has a 7--brane
configuration of the form $A^{n}BC^2$; the subalgebra is $su(n)
\otimes u(1) \otimes su(2)$ which is enhanced to a $E_n$ algebra.
It is the maximal subalgebras that we are most interested in since they obviously encode 
non--perturbative features and point out where BPS states should be in the heterotic spectrum. 

\subsection{The Orientifold Limit}
Since the 24 7--branes are non-perturbative they will not feature
directly in string models so we need to find a limit which relates them to a 
perturbative regime. A method in going between F--theory 
on K3 and Heterotic on $T^2$ is
to use Type I and I$'$ models as an intermediate step \cite{s}. From
this point of view the gauge groups in the $D_n$ series are formed by
placing $n$ D--branes on an orientifold 7--plane, ${\cal O}$. That is in
going from F--theory to Type I theory we have the 7--branes behaving as:
\begin{equation}
A^nBC \longmapsto A^n {\cal O}
\end{equation}
In collapsing the $BC$ branes to an orientifold, the NS--NS charges
cancel whilst the RR charges combine to give the correct value for
orientifold planes in eight dimensions. The maximal subalgebra
enlarges as
\begin{equation}
su(n) \otimes u(1) \longmapsto so(2n)
\end{equation}
The effect of this limit on an $E$ singularity is
\begin{equation}
A^nBC^2 \longmapsto A^n {\cal O} + C
\end{equation}
with the maximal subalgebra reorganizing itself as
\begin{equation}
su(n) \otimes u(1) \otimes su(2) \longmapsto so(2n) \otimes u(1)
\end{equation} 

Thus dual models to F--theory constructions should have enhanced gauge
groups built from these maximal subgroups. The extra states should also
be BPS. A corollary is that the corresponding Heterotic string we are
going to be interested in is HSO since it is this theory which is
S--dual to Type I \cite{pw}. 

In the following we will use HSO to denote
the heterotic string compactified on the Narain lattice $\Gamma_{2,2}
\otimes \Gamma_{16}$ while HE8 denotes $\Gamma_{2,2} \otimes
\Gamma_{8}\otimes \Gamma_{8}$. Though they are the same theories on
compactification, they have different Wilson lines when it comes to
embedding other gauge groups.

\section{Heterotic String on $T^2$}
We now turn to  building heterotic string models. From duality there are conditions to be satisfied; as 
already pointed out there can be no supersymmetry breaking. The moduli space is equivalent to that of 
compactification on a Narain lattice, prompting the restriction to gauge preserving compactifications, i.e. 
total rank is 18, and switching off background anti--symmetric tensor fields. We will assume that all 
rank 18 gauge groups appearing on the F--theory side are acceptable, i.e. have a heterotic dual; and that 
we are embedding our Wilson Lines in a lattice of the form $\Gamma_{2,2} \otimes \Gamma_{16}$ as 
opposed to $\Gamma_{2,2} \otimes \Gamma^2_{8}$; a priori this is due to the $D_n$ structure of the 
maximal subalgebra in the orientifold limit.

We compactify on the two dimensional torus \cite{ginsp}
\begin{equation}
T^2 = {\bf R}^2/2\pi\Lambda
\end{equation}
where $\Lambda$ is a lattice with basis vectors $\underline{e}_i$, $|e_i| ={1 \over R_i}$ for $i=1,2$ 
where $R_i$ are the radii of the circles. Generically $R_1 \neq R_2$. Winding number is denoted 
$\underline{\omega}_i=n^i \underline{e}_i$, $n^i \in {\bf Z}$; while momentum is given by 
$\underline{p}_i=m_i \underline{e}^{\star i}$ where $\underline{e}^{\star i}$ is a basis vector of the 
dual lattice $\Lambda^\star$. In the lattice frame, the background gauge fields are 
$\underline{A}^I_\mu= a^I_i(\underline{e}^{\star i})_\mu$ with $I=1, \ldots 16$ labelling coordinates 
in $\Gamma_{16}$ and $\mu$ the spacetime dimensions. $V$ is a vector in $\Gamma_{16}$. The 
momentum, $({\bf p_L};{\bf p_R})$ defined as
\begin{eqnarray}
{\bf p_L}&= &(V + \underline{A} \cdot \underline{\omega},\; {1 \over 2}\underline{p} - {1 \over 
2}V^K\underline{A}^K - {1 \over 4}\underline{A}^K(\underline{A}^K \cdot \underline{\omega}) + 
\underline{\omega})\\
{\bf p_R}&= &({1 \over 2}\underline{p} - {1 \over 2}V^K\underline{A}^K - {1 \over 
4}\underline{A}^K(\underline{A}^K \cdot \underline{\omega}) - \underline{\omega})
\end{eqnarray}
 form a self--dual Lorentzian lattice. The mass of a state is given by
\begin{equation}
{1 \over 4}M^2 = (N_L + {1 \over 2}{\bf p_L}^2 - 1)+(N_R + {1 \over 2}{\bf p_R}^2 - c)
\end{equation}
$N_L, N_R$ are the left and right moving oscillator numbers and $c=0, {1 \over 2}$ depending on the 
periodicity of the right moving fermions. The level matching condition is 
\begin{equation}
N_L + {1 \over 2}{\bf p_L}^2 - 1=N_R + {1 \over 2}{\bf p_R}^2 - c
\end{equation} 
Applying this to equation (9) and then imposing the condition $N_R=c$ gives the mass formula for BPS 
states
\begin{equation}
{1 \over 4}M^2 = {\bf p_R}^2
\end{equation} 
The massless vectors belonging to the roots of the underlying gauge group have $N_L=0$ along with 
${\bf p_R}^2=0, {\bf p_L}^2=2$. When the winding number is zero this gives the subgroup of the 
$SO(32)$ surviving breaking by the Wilson lines. However, for certain values of $R_i$ then further 
massless gauge bosons can appear so as to enhance the gauge group. Writing out ${\bf p_R}$ in 
component form we get
\begin{equation}
{\bf p_R} =\underline{e}^{\star i}({1 \over 2}m_i-{1 \over 2}V^Ka^K_i-{1 \over 4}a^K_ia^K_jn^j) 
- n^i\underline{e}_i
\end{equation} 
Note, that $i$ is now a label and not a component as far as the $a^K_i$ are concerned. The third term in 
the expansion looks problematic as it has the potential to cause coupling between the Wilson lines. 
However, our choices of values for the $a^K_i$ will actually give zero for the expression $a^K_ia^K_j, i 
\neq j$ and allow us to decouple the two cases. With this choice 
\begin{equation}
{\bf p_R} = \underline{e}^{\star i}({1 \over 2}m_i-{1 \over 2}V^Ka^K_i-{1 \over 4}(a_i)^2n^i) - 
n^i\underline{e}_i
\end{equation} 
where there is no summing over $i$ and $(a_i)^2$ is the length of the shortest vector of the form $a_i 
+\lambda$, where $\lambda \in \Gamma_{16}$. If the radius of compactification is, for each $i$, 
$R^2_i=1-(a_i)^2/2$ then extra massless modes appear allowing an enhancement. What actually has 
occurred here is that the two dimensional case has been split up into to two copies of a single dimensional 
compactification. They are also automatically BPS.

There are two mechanisms of gauge enhancement: $(i)$ the standard D--brane approach \cite{cjp} of 
clustering branes on top of each other; in the above notation this means identifying coordinates within the 
bulk of the fundamental cell so that the generic group $U(1)^{18}$ becomes $G_{16} \otimes U(1)^2$ 
with the $U(1)^2$ dependent only on the structure of $\Lambda$; $(ii)$ when the relationship 
$R^2_i=1-(a_i)^2/2$ is satisfied for $(a_i)$ the shortest length of the Wilson line relative to the cluster 
of D--branes we wish to enhance. However in the Type I and I$'$ dual models the second mechanism is 
non--trivial and requires the $\chi$ string discussed in \cite{bgs}. Its position in the moduli space is 
arbitrary except for gauge enhancing points when its position satisfies the above relation relating the 
radii to the length of the Wilson lines. The $\chi$ string can be related to the string junctions as its origin 
in nine dimensions is from the presence of a $D0$--brane which can couple to $D8$--branes and 
orientifold planes. It satisfies the condition that the number of Neumann--Dirichlet boundaries on the 
string is eight, ie $ND=8$ \cite{bgl1}. When we compact down to eight dimensions the $D0-D8$ 
system becomes $D1-D7$ which still satisfies $ND=8$ and is similar to the string junction system used 
in the F--theory duals.

In the work of \cite{bgs} an investigation of the $D0-D8$ system was made in nine dimensions where 
they started off with an arbitrary number, $n$ of $D0$--branes in the presence of $D8$--branes and 
$O8$--planes. It was then shown $n=1$ was required for stable configurations as in gauge enhancement. 
When there is further compactification down to eight dimensions we have $n=2$ in the decoupled 
case\footnote{It remains to be verified if this will still be the case when the Wilson lines are coupled.}. 
Decoupling basically allows us to take two copies of the nine dimensional case since we can treat the axes 
as independent except at the non--trivial point of the origin where they intersect which is only 
significant if there are branes placed at that point. 

In taking the orientifold limit of the $E_n$ series there was a C--brane ``left over''. Nevertheless, it 
contributes states necessary for the gauge enhancement and does so in a manner analogous to the states 
contributed by the $\chi$ string. We now make the tentative identification that the string junctions states 
related to the C--branes are dual to the states due to the $\chi$ string and hence the C--brane is itself 
dual to a $D1-D7$--brane set up in Type I theory (T--dual to a $D0-D8$--brane set up in nine 
dimensions). Note, that there appears to be a choice between which C--brane we should identify with the 
orientifold and which with the $\chi$ string. However, in the $D0-D8$ set up gauge enhancement occurs 
when the $D0$--brane is attached to an orientifold, and likewise here the ``left over'' C--brane is still 
at the position of the associated orientifold so it is not possible do separate their overall effects in this 
picture and the choice does not have to be made. We will return to the C--brane later when we discuss 
NS branes in heterotic theory.

For gauge groups of rank 18 there are only a finite number of ways of combining the $E_n$ groups for 
$n \geq 6$. The only one with three exceptional groups is $E_6^3$ which has been handled already in 
\cite{kst}. It also does not satisfy the decoupling feature but we will return to it later. The rest of the 
possibilities we can cluster together as $E_N \otimes E_M \otimes {\cal G}$ or $E_N \otimes {\cal G}$  
where ${\cal G}$ is of sufficient rank to make the total 18 \footnote{In the former case the rank of 
${\cal G}$ will always be less than or equal to 6. For it equal to 6 we ignore the possibility it could be 
$E_6$}. For simplicity we make it a $SO(2n)$ Lie group with no further breaking.

Looking at the first case with two exceptional groups, we can make the decomposition in the orientifold 
limit:
\begin{equation}
E_{n+1} \otimes E_{m+1} \otimes D_{16-n-m} \longmapsto D_n \otimes D_m \otimes U(1)^2 
\otimes D_{16-n-m}
\end{equation}
We can now see that we can associate the two $U(1)$ components to the two circles making up the 
compactification torus, each dimension being made responsible for the enhancement of a particular 
$D_n$ or $D_m$. The single dimensional case has already been dealt with in \cite{bgl}. For the sake of 
convenience we associate the $D_n$ group with $i=1$ and $D_m$ with $i=2$. Then we can give the 
Wilson lines as
\begin{eqnarray} 
a_1&=({1 \over 2}^n\:0^m \: 0^{16-n-m}) \nonumber \\
a_2&=(0^n \: {1 \over 2}^m \: 0^{16-n-m})
\end{eqnarray}
These trivially satisfy the condition that they decouple the $(p_R)_i$ as their product is always zero. 
Generalizations to ${\cal G} \neq D_{16-n-m}$ are straightforward. 

When the gauge group is of the form $E_N \otimes {\cal G}$ then one merely has to move the 
appropriate radius away from the critical point of enhancement in the previous case or alter one of the 
$a_i$ depending on the form desired for ${\cal G}$. The other cases of particular interest with regard to 
enhancement, $SO(36)$ and $SU(19)$ follow as in the one dimensional case with one of the Wilson 
lines set entirely to zero.

\subsection{Coupled Solutions}
We can use the duality of HSO with Type I to learn more about the structure of the moduli space of 
heterotic Wilson lines. First examine the the group $E_6 \otimes E_6 \otimes E_6$.
 
This is somewhat anomalous as $E_6 \mapsto D_5 \otimes U(1)$ requires three $U(1)$'s. The solution is 
of the form given in equation (15) with $n=m=5$. However, in order to get the third $U(1)$ for the 
enhancement the Wilson lines have the components $a^{16}_1=a^{16}_2=- {1 \over 2}$ \cite{kst}. 
This violates the decoupling argument above but provides a solution nevertheless. Hence there exist other 
solutions where the Wilson lines do not decouple. 

This is the generic case though the $E_6^3$ one is the only one with enhancing to an exceptional group 
that cannot be made to decouple. In this notation the Wilson lines act as the coordinates of a square 
moduli space of axes $a_1, a_2$ such that $0 \leq a_i \leq {1 \over 2}$. Each pair of components of the 
Wilson lines, $(a_1, a_2)^K$ now forms the coordinate of the Kth D--brane when it is mapped to a dual 
Type I model in eight dimensions. The orientifold planes are represented by the corners $(0,0),\:(0,{1 
\over 2}),\:({1 \over 2},0),\:({1 \over 2},{1 \over 2})$ though they only have an effect if there are 
D--branes on them; decoupled solutions lie purely on the axes. However, this space is only a 
fundamental cell of a larger lattice and extra massless states can arise, as in $E_6^3$, when D-branes are 
located at special points outside the fundamental cell when other winding modes become massless. These 
situations will break the $Z_4$ symmetry of the Wilson lines that exist when the D--branes all lie 
within the fundamental cell.

Coupled solutions lie within the bulk and represent the relative difficulty of finding the solution as the 
positions here are arbitrary, the solutions giving rise to gauge groups lying on loci as opposed to 
particular points.  For many cases the loci of solutions will intersect with the boundary and the decoupled 
form of the Wilson lines can be recovered.

A final set of gauge groups of interest are those of the form $D_n^x \otimes D_m^y$ with $nx+my=18$. 
It can be shown that if $x+y \geq 4$ then there will always be more than 24 branes required on the 
F--theory side. That is if $x+y > 4$ then some of the gauge groups would have to have rank less than 4 
and thus are in the $A$ series as opposed to the $D$, in line with the fact that in the Type I picture there 
are only four orientifold planes. In the case $x+y = 4$ then the gauge group is $D_4^2 \otimes D_5^2$ 
which from the F--theory side is not acceptable as it would require 26 7--branes. On the HSO side it 
would be constructed by placing 4 D--branes on each orientifold plane and choosing the appropriate 
radii to enhance two of the $D_4$ to $E_5$ which would give us back the $D_5$ gauge groups. The 
Wilson lines are:
\begin{eqnarray}
a_1&=(0^4 \: 0^4 \: {1 \over 2}^4 \: {1 \over 2}^4)  \nonumber \\
a_2&=(0^4 \: {1 \over 2}^4\:0^4 \: {1 \over 2}^4)
\end{eqnarray}
The resolution lies in the fact that is not possible to single out only two $D_4$ we want to enhance and 
the construction from the heterotic point of view breaks down as well. There are no other cases such as 
this so we are justified in the assumption that all rank 18 gauge groups which can be constructed from 
F--theory on K3 have an appropriate heterotic dual.

So far we have been concentrating purely on the HSO type models. In theory we should be able to 
construct a dictionary for embedding in HE8 since it has the same moduli space and is related to HSO by 
T--duality. However, there is no simple orientifold limit as for HSO and the Wilson lines are 
non--trivial. A case in hand is $E^3_6$ which, to build up in $\Gamma_{2,2} \otimes \Gamma^2_{8}$ 
the third $E_6$, we require a $su(3)^3$ maximal subalgebra, the brane structure for which is not 
apparent. The generic $SL(2, {\bf Z})$ transformation to convert the standard 
$A^{n_a}B^{n_b}C^{n_c}$ configuration to a form where the maximal subalgebra reflects the 
embedding in $E_{8} \otimes E_{8}$ has not been constructed yet. If the T--duality is modified as 
discussed in the next section, then it may be the case that this transformation does not exist.

\section{Comparison with NS9--Branes and Conclusion}

In recent papers Hull \cite{h} has discussed the existence of NS9-branes in non--perturbative HSO theory. NS9--branes are the S--duals of the D9--branes in Type I theory and can also be deduced from M9-branes in M--theory \cite{b}. Their discovery, implied by duality, gives the same brane structure in the heterotic string that has proved to be so rich in Type I theory. In particular it is not hard to see that the heterotic Wilson lines should now correspond to the position of the 16 NS9--branes. This in turn provides evidence that the map between the Wilson lines of Type I and HSO is exact under S--duality; the RR charge of the orientifolds will change to NS--NS in HSO models. 

In deriving the map between F--theory on K3 and HSO on $T^2$ we have implicitly assumed that the A--branes of F--theory have the same charge as $D7$--branes. This is not strictly necessary as what mattered in the above construction was the correct cancellation and summing of overall charge along with the maximal subalgebra. This is an intrinsic feature of the F--theory construction as various brane configurations are considered equivalent if they can be related by an $SL(2, {\bf Z})$ transformation 
\cite{gz}. The $SL(2, {\bf Z})$ self--duality of the parent Type IIB theory includes an S--duality. Thus the standard $A^{n_a}B^{n_b}C^{n_c}$ configuration can be related to another configuration of charged 7--branes, $\tilde{A}^{n_a} \tilde{B} ^{n_b} \tilde{C} ^{n_c}$, so that the gauge structure is exactly the same but the NS--NS and RR charges are interchanged. Hence, when examining the perturbative string models we can only tell the difference between HSO or Type I from the charges of the F--theory configuration we started with. In terms of the moduli space, it provides evidence that HSO and Type I with all their permitted Wilson line configurations are equivalent up to gauge group/Kodaira classification, and that the subgroup structure discussed above persists under an S--duality transformation. This ties in nicely with the use of truncation techniques on the parent Type IIB spectrum in ten dimensions to obtain the Type I and HSO theories performed in \cite{Bergshoeff:1999bx}

In this paper we have constructed an explicit map taking us from the Kodaira classification of singularities in F--theory compactified on elliptically fibered K3 surfaces to the Wilson lines in the Heterotic $SO(32)$ string compactified on $T^2$, and discussed some issues arising out of gauge enhancement. 
\newline
{\bf Acknowledgments.} We would like to extend our thanks to D.C. Dunbar and M.R. Gaberdiel for explanation of their work. We are also grateful to P. Aspinwall, L. Bonora, M. Gross, C. Johnson, H. Skarke and Swansea theory group for a series of useful conversions and communications. This work was supported by PPARC. 
\newline
As this manuscript was in preparation ref \cite{i} appeared, the results of which overlap with this paper.

\end{document}